\documentstyle[12pt]{article}
\def\beq{\begin{equation}}
\def\eeq{\end{equation}}
\def\bea{\begin{eqnarray}}
\def\eea{\end{eqnarray}}

\def\ba{\begin{array}}
\def\ea{\end{array}}
\def\ds{\displaystyle}

\def\,{\"{U}}
\def\6{\.{I}}

\begin{document}

\title{Effective Mass Schr\"{o}dinger Equation for Exactly 
Solvable Class of One-Dimensional Potentials}

\author{Metin Akta\c{s}$^a$, Ramazan Sever$^b$\thanks{Corresponding author:
 E-mail: sever@metu.edu.tr}\\[.5cm]
$^{a}$Department of Physics, Faculty of Arts and Sciences\\
K{\i}r{\i}kkale University, 71450, K{\i}r{\i}kkale, Turkey\\[.5cm]
$^{b}$Department of Physics, Middle East Technical University\\
06531 Ankara, Turkey}

\date{\today}
\maketitle
\normalsize
\begin{abstract}

\noindent We deal with the exact solutions of Schr\"{o}dinger
equation characterized by position-dependent effective mass via
point canonical transformations. The Morse, P\"{o}schl-Teller and
Hulth\'{e}n type potentials are considered respectively. With the
choice of position-dependent mass forms, exactly solvable target
potentials are constructed. Their energy of the bound states and
corresponding wavefunctions are determined exactly.\\

\end{abstract}

\newpage
\section{Introduction}
\noindent There has recently been a continual interest to the
solutions of Schr\"{o}dinger equations describing systems with
spatially dependent effective masses [1-11]. Systems with certain
types of effective masses are found to be useful in the
determination of physical properties of semiconductor
heterostructures [12], quantum liquids and dots [13, 14], helium and
metal type clusters [15, 16] and nuclei [17] as well. In addition to
these, the quantum mechanics for systems of particle with some types
of masses has not only been considered for the instantaneous
Galilean invariance [18], but also studied for the path integral
approach [19] and $su(1,1)$ Lie algebraic technique etc. [20, 21].
No matter how difficult exact solutions of are to obtain, the
effective mass Schr\"{o}dinger equation for certain type potentials
can be solved [22, 23]. Within the framework of SUSYQM, problems
including the shape invariance requirement are solved exactly
[24-27]. In the mapping of the nonconstant mass Schr\"{o}dinger
equation, point canonical transformations (PCTs) are employed
[28-31]. When dealing with this process, we need to transform it to
a constant mass one so that the latter equation can be solved
easily. Therefore, it will yield both the energy spectra and
corresponding wavefunctions of the target problem with regard to
those of reference problem. Some reference (original) potentials
satisfying the concept of exactly solvability such as oscillator,
Coulomb, Morse [1], hard-core potential [32], trigonometric type
[33] and conditionally exactly solvable potentials [2, 31] as well
as the Scarf and Rosen-Morse type [34] ones including the
PT-symmetry are considered for the construction of exact solution
via PCT. In the present work, PCT approach is applied to the exact
solutions of the nonconstant mass Schr\"{o}dinger equation for
Morse, P\"{o}schl-Teller and Hulth\'{e}n potentials respectively.

\noindent The plan of the article is in the following. In Sec. II,
we show how to map the effective mass Schr\"{o}dinger equation by
using point canonical transformation method. In Sec. III, we apply
the procedure to construct the target problem, including the energy
spectrum and corresponding wavefunctions, for potentials mentioned
above. Finally, we draw some remarkable conclusions in Sec. IV.

\section{Mapping of the Effective Mass Schr\"{o}dinger Equation}
\noindent The one dimensional time independent Schr\"{o}dinger
equation within the case of spatially dependent mass is written

\begin{equation}\label{2}
    -\frac{1}{2}\left[\nabla_{x}~\frac{1}{M(x)}~\nabla_{x}\right]\Psi(x)
-[E-V(x)]\Psi(x)=0,
\end{equation}

\bigskip
\noindent where $M(x)=m_0m(x)$. More explicit form of this equation
is

\begin{equation}\label{3}
    \Psi^{\prime\prime}(x)-\left(\frac{m^{\prime}}{m}\right)\Psi^{\prime}(x)+2m[E-V(x)]\Psi(x)=0,
\end{equation}

\noindent where we set $\hbar=1$ and $m_0$ as constant, prime and
double prime factors indicate the first and second order derivatives
with respect to $x$. However, the one dimensional Schr\"{o}dinger
equation with a constant mass can be written as

\begin{equation}\label{4}
    \Phi^{\prime\prime}(y)+2[\varepsilon-V(y)]\Phi(y)=0.
\end{equation}

\noindent By introducing a transformation $y\rightarrow x$ through a
mapping function $y=f(x)$ and rewriting the wavefunction

\begin{equation}\label{5}
    \Phi(y)=g(x)\Psi(x),
\end{equation}

\noindent Schr\"{o}dinger equation with constant mass is transformed
to
\smallskip

\begin{equation}\label{6}
    \Psi^{\prime\prime}(x)+2\left(\frac{g^{\prime}}{g}-\frac{f^{\prime\prime}}{2f^{\prime}}\right)\Psi^{\prime}(x)
    +\left\{\left(\frac{g^{\prime\prime}}{g}-\frac{f^{\prime\prime}}{f^{\prime}}\frac{g^{\prime}}{g}\right)
    +2(f^{\prime})^{2}[\varepsilon-V(f(x))]\right\}\Psi(x)=0.
\end{equation}

\smallskip
\noindent By comparing each sides of the Eqs. (2) and (5)
term-by-term, we can identify the following conditions

\begin{equation}\label{7}
    g(x)=\left(\frac{f^{\prime}(x)}{m(x)}\right)^{1/2}
\end{equation}

\noindent and

\begin{equation}\label{8}
    E-V(x)=\frac{(f^{\prime})^{2}}{m}\left[\varepsilon-V(f(x))\right]
+\frac{1}{2m}F(f,g),
\end{equation}

\bigskip
\noindent where
$\ds{F(f,g)=\left(\frac{g^{\prime\prime}}{g}-\frac{f^{\prime\prime}}{f^{\prime}}\frac{g^{\prime}}{g}\right)}$.
If we consider the substitution map as $(f^{\prime})^{2}=m$ in Eq.
(7), our reference problem is transformed to the target problem
including the energy spectra of the bound states, potential and
wavefunction as

\begin{eqnarray}\label{9}
  E_{n} &=& \varepsilon_{n}\nonumber \\[0.2cm]
  V(x) &=&V(f(x))+\frac{1}{8m}\left[\frac{m^{\prime\prime}}{m}-\frac{7}{4}\left(\frac{m^{\prime}}{m}\right)^{2}\right]\\[0.2cm]
   \Psi_{n}(x)&=& [m(x)]^{1/4}~\Phi_{n}(f(x))\nonumber.
\end{eqnarray}

\noindent Here, we point out that if we have a problem whose exact
solution is well established, then we can apply the $PCT$ method to
this problem. That is, it preserves the structure of the wave
equation of the target problem that has the same class as that of
the reference problem.

\section{Applications}

\noindent In this section, we deal with three different
position-dependent mass distributions. The reference potentials such
as Morse, P\"{o}schl-Teller and Hulth\'{e}n [35] are taken. Then, we
are going to construct their target systems respectively.

\subsection{Asymptotically vanishing mass distribution: $m(x)=\frac{\alpha^2}{x^{2}+q}$}

\noindent The mapping function is

\begin{equation}\label{11}
    y=f(x)=\int^x \sqrt{m(x)}~dx=\alpha\ell n\left(x+\sqrt{x^{2}+q}\right),
\end{equation}

\noindent with
\smallskip

\begin{equation}\label{12}
    x=sinh_q\left(\frac{y}{\alpha}\right),
\end{equation}

\noindent where $\alpha\neq 0$.

\subsubsection{Morse Case}
\noindent Let us first consider the Morse potential as the reference
problem [35, 36]

\begin{equation}\label{15}
    V(y)=D(e^{-2\alpha y}-2e^{-\alpha y})\equiv D\left[(1-e^{-\alpha
    y})^{2}-1\right].
\end{equation}

\noindent This is the source potential with the energy eigenvalues
and eigenfunctions as [35]

\begin{eqnarray}\label{16}
  \varepsilon_{n} &=&-\frac{\alpha^{2}\hbar^{2}}{2m} \left[\bar{D}-\left(n+\frac{1}{2}\right)\right]^{2}
  \nonumber\\[0.3cm]
  \Phi_{n}(y) &=& C_{n}z^{\beta}e^{-\eta z}L_{n}^{t}(z),
\end{eqnarray}

\noindent with $z=-e^{\alpha y}$. By putting the mass function and
Eq (9) into the Eq. (8) and using (10), the target problem can be
constructed as follows:

\smallskip
\begin{eqnarray}\label{17}
  E_{n}&=& \varepsilon_{n} \nonumber\\[0.2cm]
  V(x) &=& D\left\{\left[1+\alpha^{2}\left(x+\sqrt{x^{2}+q}\right)\right]^{2}-1\right\}
  +\frac{1}{8\alpha^{2}}\left[1+\left(\frac{q}{x^{2}+q}\right)\right]
  \\[0.2cm]
  \Psi_{n}(x) &=& C_{\bar{n}}(x^{2}+q)^{-1/4}[f(x)]^{\beta}e^{-\eta
  f(x)}L_{n}^{t}(f(x))\nonumber,
\end{eqnarray}

\noindent where $C_{\bar{n}}=\sqrt{\alpha}C_{n}$.

\subsubsection{P\"{o}schl-Teller Case}
\noindent Now, we will consider the potential as [35, 36]
\smallskip
\begin{equation}\label{20}
    V(y)=-\frac{U_{0}}{\cosh^{2}(\alpha y)}\equiv \frac{-4U_0}{(e^{\alpha y}+e^{-\alpha
    y})^{2}},\qquad
    U_{0}=\lambda(\lambda-1).
\end{equation}
\smallskip
\noindent Our reference potential is now order that its energy
spectra and wavefunction are given [35]

\smallskip
\begin{eqnarray}\label{21}
  \bar{\varepsilon}_{n} &=&
  A^{2}-\frac{\alpha^{}\hbar^{2}}{2m}\left[-\left(n+\frac{1}{2}\right)+\frac{1}{2}\sqrt{1+4\gamma^{2}}\right]\nonumber\\[0.2cm]
  \Phi_{n}(y) &=& C_{n}(1-z^{2})^{\beta/2}P_{n}^{(\beta,~\beta)}(z),
\end{eqnarray}

\noindent with $z=\tanh(\alpha y)$. In this case, we want to
construct the target problem for asymptotically vanishing mass
distribution. Following the same procedure as in above, we get the
target system

\begin{eqnarray}\label{22}
  E_{n}&=& \bar{\varepsilon}_{n} \nonumber\\[0.2cm]
  V(x) &=& -\frac{U_{0}}{x^{2}+q}+\frac{1}{8\alpha^{2}}\left[1+\left(\frac{q}{x^{2}+q}\right)\right]\\[0.2cm]
  \Psi_{n}(x) &=& C_{\bar{n}}(x^{2}+q)^{-1/4}[1-f^{2}(x)]^{\beta/2}P_{n}^{(\beta,~\beta)}(f(x))\nonumber,
\end{eqnarray}

\noindent where $C_{\bar{n}}=\sqrt{\alpha}C_{n}$.

\subsubsection{Hulth\'{e}n Case}

\noindent Hulth\'{e}n potential is defined [35, 36]
\begin{equation}\label{25}
    V(y)=-V_{0}\frac{e^{-\alpha y}}{(1-e^{-\alpha
    y})}\equiv -V_{0}(e^{\alpha y}-1)^{-1}.
\end{equation}

\noindent Its energy spectrum and corresponding wavefunctions are
\smallskip
\begin{eqnarray}\label{26}
  \tilde{\varepsilon}_{n} &=&-V_{0}\left[\frac{\beta^{2}-\bar{n}^{2}}{2\bar{n}\beta}\right]^{2}\nonumber\\[0.2cm]
  \Phi_{n}(y) &=& C_{n}z^{\tilde{\varepsilon}}(1-z)^{\mu/2}P_{n}^{(2\tilde{\varepsilon},~\mu-1)}(1-2z),
\end{eqnarray}

\noindent with $z=e^{-\alpha y}$ [35]. The target system is
constructed with the mass function

\smallskip
\begin{eqnarray}\label{27}
  E_{n}&=& \tilde{\varepsilon}_{n} \nonumber\\[0.2cm]
  V(x) &=& -V_{0}\left[-1+\alpha^{2}\left(x+\sqrt{x^{2}+q}\right)\right]^{-1}
  +\frac{1}{8\alpha^{2}}\left[1+\left(\frac{q}{x^{2}+q}\right)\right]\\[0.2cm]
  \Psi_{n}(x) &=& C_{\bar{n}}(x^{2}+q)^{-1/4}[f(x)]^{\tilde{\varepsilon}}[1-f(x)]^{\mu/2}
  P_{n}^{(2\tilde{\varepsilon},~\mu-1)}(1-2f(x))\nonumber.
\end{eqnarray}

\subsection{Hyperbolic Mass Distribution I: $m(x)=\tanh_{q}^{2}(\alpha x)$}

\noindent As a second example, we want to deal with the square
hyperbolic mass functions. The resulting map for the mass type
becomes

\begin{equation}
y=\bar{f}(x)=\frac{1}{\alpha}\ell n\cosh_{q}(\alpha x)
\end{equation}

\noindent with $x=\frac{1}{\alpha}\cosh_{q}^{-1}(e^{\alpha y})$.

\subsubsection{Morse Case}

\noindent For the case, the potential functions and the
wavefunctions having the same spectrum are obtained as
\smallskip
\begin{eqnarray}\label{18}
V(x)&=& D\left\{\left[1+\cosh_{q}(\alpha
x)\right]^{2}-1\right\}-\frac{\alpha^{2}/2}{\sinh_{q}^{4}(\alpha
x)}\left[\frac{5}{4}+\sinh_{q}^{2}(\alpha
x)\right]\nonumber\\[0.2cm]
\Psi_{n}(x)&=&C_{\bar{n}}\sqrt{[\tanh_{q}(\alpha
x)]}[\bar{f}(x)]^{\beta}e^{-\eta\bar{f}(x)}L_{n}^{t}(\bar{f}(x)),
\end{eqnarray}

\noindent where the hyperbolic functions are $cosh_{q}(x)=(e^{x}+q
e^{-x})/2$, $sinh_{q}(x)=(e^{x}-q e^{-x})/2$ and
$tanh_{q}(x)=\left(\frac{sinh_{q}(x)}{cosh_{q}(x)}\right)$ [37].

\subsubsection{P\"{o}schl-Teller Case}

\noindent By following the $PCT$ procedure, the target system is
constructed

\begin{eqnarray}\label{23}
V(x)&=& -4U_{0}\left[\cosh_{q}(\alpha x)+sech_q(\alpha
x)\right]^{-2}-\frac{\alpha^{2}/2}{\sinh_{q}^{4}(\alpha x)}\left[\frac{5}{4}+\sinh_{q}^{2}(\alpha x)\right]\nonumber\\[0.2cm]
\Psi_{n}(x)&=&C_{\bar{n}}\sqrt{[\tanh_{q}(\alpha
x)]}[1-(\bar{f}(x))^2]^{\beta/2}P_{n}^{(\beta,~\beta)}(\bar{f}(x)),
\end{eqnarray}

\noindent where $sech_{q}(x)=\frac{1}{cosh_{q}(x)}$ [37]. Hence, the
energy spectrum of the bound state is the same as that of the
reference one.

\subsubsection{Hulth\'{e}n Case}

\noindent The mapping function leads to the following target system

\begin{eqnarray}\label{28}
  V(x) &=& -V_{0}\left[-1+\cosh_{q}(\alpha x)\right]^{-1}
  -\frac{\alpha^{2}/2}{\sinh_{q}^{4}(\alpha x)}\left[\frac{5}{4}+\sinh_{q}^{2}(\alpha
  x)\nonumber\right]\\[0.2cm]
   \Psi_{n}(x)&=&C_{\bar{n}}\sqrt{[\tanh_{q}(\alpha x)]}[\bar{f}(x)]^{\tilde{\varepsilon}}[1-\bar{f}(x)]^{\mu/2}
  P_{n}^{(2\tilde{\varepsilon},~\mu-1)}(1-2\bar{f}(x)).
\end{eqnarray}

\noindent This system and its original one share the same spectrum.

\subsection{Hyperbolic Mass Distribution II: $m(x)=\coth_{q}^{2}(\alpha x)$}

\noindent If we consider the second type hyperbolic mass function,
we also get

\begin{equation}
y=\tilde{f}(x)=\frac{1}{\alpha}\ell n\sinh_{q}(\alpha x),
\end{equation}

\noindent with $x=\frac{1}{\alpha}\sinh_{q}^{-1}(e^{\alpha y})$.

\subsubsection{Morse Case}

\noindent This mapping yields the potential function and the
wavefunction with the same spectrum

\begin{eqnarray}\label{19}
V(x)&=& D \left\{\left[1+\sinh_{q}(\alpha
x)\right]^{2}-1\right\}-\frac{\alpha^{2}}{2}\left[\frac{1}{\sinh_{q}^{2}(\alpha
x)}+\frac{(9/4)}{\cosh_{q}^{4}(\alpha x)}\right]\nonumber\\[0.2cm]
\Psi_{n}(x)&=&C_{\bar{n}}\sqrt{[\coth_{q}(\alpha
x)]}[\tilde{f}(x)]^{\beta}e^{-\eta\tilde{f}(x)}L_{n}^{t}(\tilde{f}(x)),
\end{eqnarray}

\noindent where
$coth_{q}(x)=\left(\frac{cosh_{q}(x)}{sinh_{q}(x)}\right)$ [37].

\subsubsection{P\"{o}schl-Teller Case}

\noindent The target system with the same spectrum is then

\begin{eqnarray}\label{24}
V(x)&=& -4U_{0}[\sinh_{q}(\alpha x)+cosech_q(\alpha
x)]^{-2}-\frac{\alpha^{2}}{2}\left[\frac{1}{\sinh_{q}^{2}(\alpha
x)}+\frac{(9/4)}{\cosh_{q}^{4}(\alpha x)}\right]\nonumber\\[0.2cm]
\Psi_{n}(x)&=&C_{\bar{n}}\sqrt{[\coth_{q}(\alpha
x)]}[1-(\tilde{f}(x))^2]^{\beta/2}P_{n}^{(\beta,~\beta)}(\tilde{f}(x)),
\end{eqnarray}

\noindent where $C_{\bar{n}}=\sqrt{\alpha}C_{n}$ and
$cosech_{q}(x)=\frac{1}{sinh_{q}(x)}$ [37].

\subsubsection{Hulth\'{e}n Case}

\noindent Finally, we construct the target system with the same
energy spectrum of the bound state

\begin{eqnarray}\label{29}
  V(x) &=& -V_{0}\left[-1+\sinh_{q}(\alpha x)\right]^{-1}
  -\frac{\alpha^{2}}{2}\left[\frac{1}{\sinh_{q}^{2}(\alpha
x)}+\frac{(9/4)}{\cosh_{q}^{4}(\alpha x)}\right]\nonumber\\[0.2cm]
   \Psi_{n}(x)&=&C_{\bar{n}}\sqrt{[\coth_{q}(\alpha x)]}[\tilde{f}(x)]^{\tilde{\varepsilon}}[1-\tilde{f}(x)]^{\mu/2}
  P_{n}^{(2\tilde{\varepsilon},~\mu-1)}(1-2\tilde{f}(x)),
\end{eqnarray}

\noindent with $C_{\bar{n}}=\sqrt{\alpha}C_{n}$.

\section{Conclusions}
\noindent In the present work we have applied the $PCT$ approach for
systems with some spatially dependent effective masses, exactly
solvable potentials such as Morse, P\"{o}schl-Teller and
Hulth\'{e}n, whose bound-state energies and corresponding
wavefunctions are determined algebraically. The determination of the
mapping function plays an important role on the construction of the
target system which has the exact closed forms of the energy
spectrum and corresponding wavefunctions. In particular, the former
system with the source potential and the latter one with the target
potential will share the same bound-state spectra. In this paper we
use the deformation parameter $q$ describing the mass distributions.
For instance, when $q=1$, the target potentials and wavefunctions of
all potentials mentioned above are identical so that the energy
spectra of the systems comply with that of the reference system
[35]. We also point out as a final remark that both the form of the
potential and spatial dependence of the mass $m(x)$ cause a system
to exactly solvability.

\section{Acknowledgment}
This work is partially supported by the Turkish Council of
Scientific and Technological Research (TUBITAK).

\end{document}